\documentclass[prl,twocolumn,showpacs,preprintnumbers,amsmath,amssymb]{revtex4}

\usepackage{graphicx}
\usepackage{dcolumn}
\usepackage{bm}
\usepackage[latin1]{inputenc}          
\usepackage[english]{babel}

\begin{document}

\title{Ultrafast Light-Induced Coherent Optical and Acoustic Phonons in few Quintuple Layers of Topological Insulators Bi$_2$Te$_3$. }  

\author{M. Weis$^\ddagger$,$^\dagger$, K. Balin$^\ddagger$, R. Rapacz$^\ddagger$, A. Nowak$^\ddagger$, M. Lejman$^\dagger$, J. Szade$^\ddagger$\footnote{ Electronic address: jacek.szade@us.edu.pl}, P. Ruello$^\dagger$ \footnote{ Electronic address: pascal.ruello@univ-lemans.fr}}
\affiliation{
$^\ddagger$ A. Che{\l}kowski Institute of Physics and Silesian Center for Education and Interdisciplinary Research, 75 Pu{\l}ku Piechoty 1A, 41-500 Chorz\'ow, University of Silesia, Poland.\\
$^\dagger$ Institut des Mol\'ecules et Mat\'eriaux du Mans, UMR CNRS 6283, 
Universit\'e du Maine, 72085 Le Mans,  
France.}

\begin{abstract}
Ultrafast lattice dynamics of few quintuple layers of  topological insulator (TI) Bi$_2$Te$_3$ is studied with time-resolved optical pump-probe spectroscopy. Both optical and acoustic phonons are photogenerated and detected. Here, in order to get new insights on the out-of-equilibrium electron-phonon coupling and phonons dynamics in confined TI, different nanostructures have been investigated (single or polycrystalline QLs assemblies and nano-crystallized islands). Contrary to previous literature claims, we show that even for nanostructures containing only 10 quintuple layers (QLs), the symmetric A1g(I) coherent optical phonon is efficiently photogenerated and no restriction due to the structural confinement appears.  We also observe that whatever the arrangement of the nanostructures, the A1g(I) optical phonon features are similar (lifetime). We also report the observation of confined coherent acoustic phonons propagating from QLs to QLs whose spectrum is, this time, very sensitive to the atomic arrangement. In the case of the single crystalline ultrathin film, the time of flight analysis of these acoustic phonons provides direct estimate of the elastic properties of these nanostructures as well as some estimates of Van der Waals interactions between QLs.  

\end{abstract}

\pacs{63.20.kd  78.66.-w  78.47.J-   78.47.D- }
\maketitle

\section{Introduction}

Topological insulators (TI) are a new electronic state in condensed matter with semiconducting bulk solid while conducting on the surface. Moreover, the surface Dirac electronic states  exhibit natural spin-polarized current and appear to be robust regarding to backward non-magnetic scattering which is unique and heralds possible very attractive applications for next generation spintronic devices \cite{moore}. Beside the crucial spin-orbit coupling effect on the features of surface and bulk electrons dynamics  \cite{haj} and the features of confined electrons in 2D TIs \cite{zhang}, the electron-phonon coupling is the object of active discussions and plays a pivotal role in the transport properties of the TIs \cite{cos,kim,reij,stein}. This is also of prime importance for the thermoelectric properties for this class of materials. In particular, it has been observed recently in Bi$_2$Se$_3$ (BS) that the longitudinal optical phonon (LO) with an energy of 8 meV plays a crucial role in the electrons scattering and sets a limit in the conductivity \cite{cos}. This LO mode is the Raman active mode A1g(I) (2.13 THz in Bi$_2$Se$_3$ \cite{jusserand} and 1.85 THz in Bi$_2$Te$_3$\cite{dekor}). The electron-acoustic phonon coupling through the deformation potential has also an important contribution that likely limits the electrical conductivity \cite{kim}. Additionally, some theoretical reports suggest some driving contributions of this electron-phonon coupling in the appearance of the topological state \cite{cana}. Witnessing probably the large electron-phonon coupling, it is remarkable that this A1g(I) longitudinal optical phonon is the one which is efficiently photoexcited by a femtosecond laser optical pulse in bulk Bi$_2$Te$_3$\cite{dekor,kama,nori} or Bi$_2$Se$_3$\cite{kumar}. However, Wang et al \cite{wang} recently reported that the large confinement of light-induced hot electrons in Bi$_2$Te$_3$ films (BT) as thin as 10 nm down to 5 nm cannot lead to sufficient large photoinduced force to induce lattice displacement. This peculiar situation could be a drawback in the perspective of manipulation of coherent phonons in such nanostructures. Moreover, these report could indicate that hot electrons could couple differently in bulk BT than in thin films made of few BT quintuple layers (QLs shown in Fig. \ref{fig1}) which is a fundamental question as well as crucial for applications in future nanodevices. Few nanometers thick thin films are key nanostructures since they are naturally a rich playground to investigate the electron-phonon coupling with variable ratio of Dirac-like surface electrons to bulk electrons and also because, in the particular case of Bi$_2$Te$_3$ at least, the characteristic length over which the intervalley scattering process between surface and bulk electrons is effective has been recently estimated as as 5 nm which actually scales with 5 QLs\cite{haj}.

In this article, femtosecond pump-probe spectroscopy has been used to demonstrate that contrary to what has been previously proposed, it is actually possible to generate coherent optical phonons in BT nanostructures and a discussion of the driving force is developed. The mechanisms of excitation and detection are efficient whatever the nanostructure of the system is (single crystalline film, polycrystalline film or crystallized islands).  The lifetime of the A1g optical mode appears to be pretty insensitive to the long range crystallographic order showing the LO decay is a local mechanism. Additionally to optical phonons, coherent acoustic phonons have also been laser generated and detected with this technique. In particular, we report the observation of confined quantized coherent acoustic eigenmodes in the single crystalline 15 nm thick film. The analysis of the eigenmodes frequencies leads to the estimates of the sound velocity (i.e. elastic modulus) along the [001] direction, which is the direction along which the QLs are connected with Van der Waals bonds (Fig. \ref{fig1}(c)). A discussion on the electron-acoustic phonon coupling mechanisms is also given. 



\section{Samples growth and methods}

\begin{figure*}[t]
\centerline{\includegraphics[width=16cm]{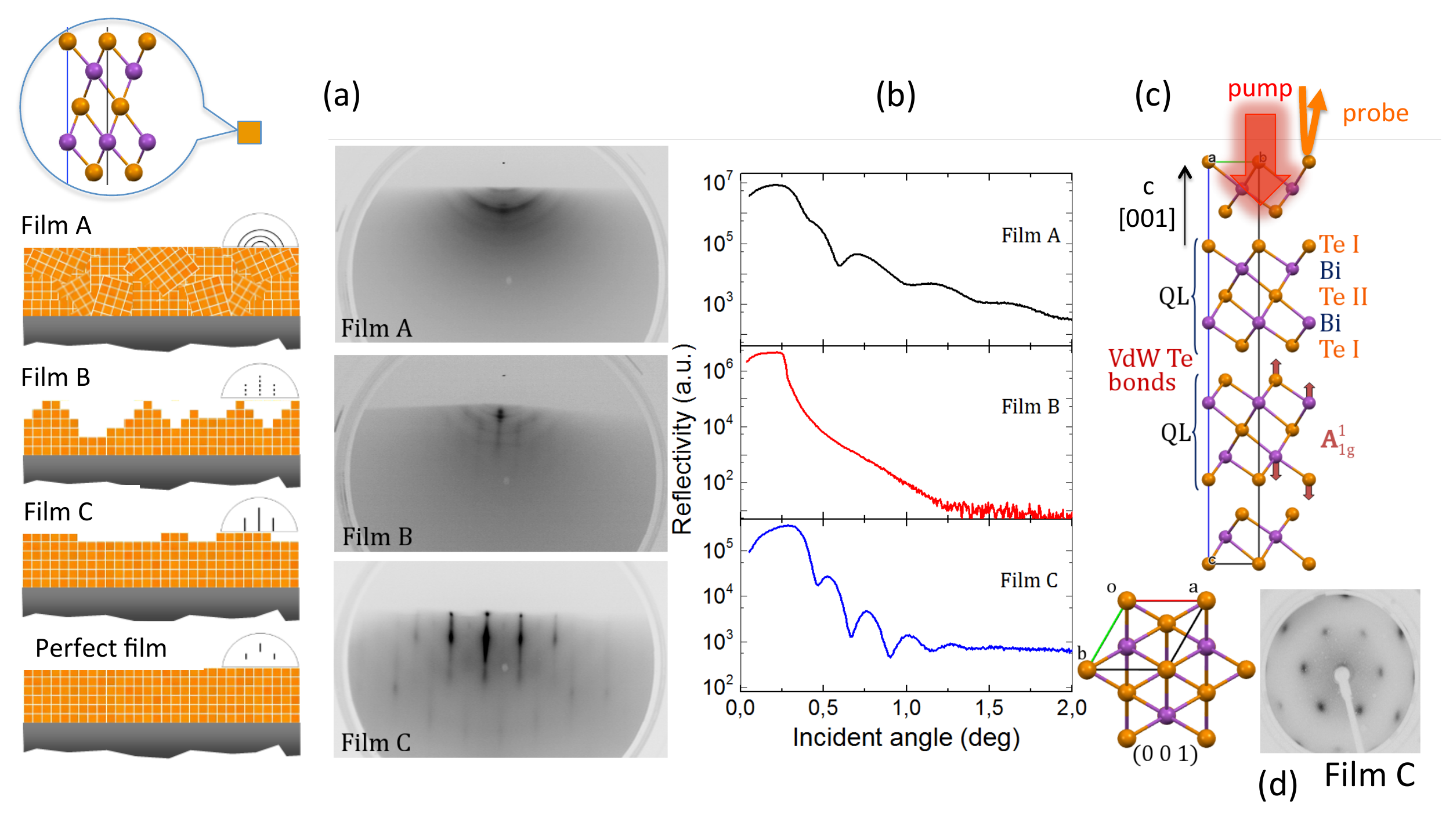}}
\caption{\label{fig1} (color online)  Physical characteristics of the thin TI films : (a) experimental RHEED images for films labelled as A, B and C (see details in the text for the growing conditions). The corresponding sketch of the nanostructures is given where QL is represented by a ball, along with the corresponding theoretical RHEED images. (b) X-ray reflectivity spectra. (c) pictures of the crystallographic structure of Bi$_2$Te$_3$ with the QLs connected through Van der Waals Te bonds.  (d) The bottom inset is a LEED image of the film C showing its single crystallinity and the 6-fold symmetry of the (001) plane.}
\end{figure*}

All studied films of Bi$_x$Te$_y$ with various compositions have been grown on Si(100) or mica (muscovite) substrates by thermal evaporation with the use of Molecular Beam Epitaxy (MBE) system from Prevac. The growth was performed in the co-deposition mode. The electronic and crystallographic characterizations were carried out in-situ with the use of x-ray photoelectron spectroscopy (XPS), the reflection high energy electron diffraction (RHEED), the low energy electron diffraction (LEED) and the atomic force microscopy (AFM). Additionally, the ex-situ x-ray reflectivity (XRR) measurements were performed. The Fig. \ref{fig1} shows the different characteristics of each BT film.
Preparation of the Si substrate and details of the films grown on Si (samples A and B) are given in our earlier work \cite{jacek}. The film C was deposited on the (110) freshly cleaved Muscovite mica substrate (Ted Pella, Inc.) which was then was annealed at 230$^\circ$C in the MBE chamber by 1.5 hour. For both types of the substrate it was kept at 130$^\circ$C during the deposition. The deposition rate has been controlled by the quartz crystal micro balance. 
The film A shows a polycrystalline structure while for the sample B the well visible streaks in the RHEED pattern indicate at least a partial long range atomic order on the surface (Fig. \ref{fig1}(a)). For that sample B we found no oscillations in the XRR results (Fig. \ref{fig1}(b)) which can be related to a large roughness detected in the AFM tests where the value of RMS up to 5-6 nm (per 1 $\mu$m$^2$) was obtained. This film is then more arranged according to some crystallized islands induced by the  growing process. It is worth to mention that the value of RMS for the substrates was lower than 1 nm. For the film C deposited on mica, the electron diffraction experiments (RHEED in Fig. \ref{fig1}(a) and LEED in Fig. \ref{fig1}(d)) show that this film is single crystalline. 
The thickness of the films was derived from the quartz crystal micro balance and XRR data. For the sample C grown on mica a good agreement was found between the data obtained from two methods (about 15 nm). For the sample A, the BT layer has a thickness close to 10 nm but fitting of the XRR data indicates the presence of an overlayer of lower density and a thickness of about 7 nm. The upper layer is formed probably by Te and tellurium oxides. The XPS analysis (not shown) allowed to determine the atomic composition and chemical state of the components. All the films appear to be free of any contamination and the atomic composition characterized by the Bi/Te ratio varied from 34/66 (film A), through 44/56 (film B) to 46/54 (film C). The analysis of the Bi and Te most prominent photoemission lines showed for films B and C only one chemical state with the positions of the characteristic lines of Bi$_2$Te$_3$ \cite{jacek}. For the sample A an additional line in the Te 3d spectrum was found which could be assigned to pure Te. The Bi$_x$Te$_y$ system is known to form many layered nanostructures especially for the compositions $y/x <$ 3/2. The stacking sequence of the quintuple layers (QL), characteristic for Bi$_2$Te$_3$ and additional layers of Bi leads to formation of many structures \cite{bos}. 

The pump probe technique used here is based on a 80 MHz repetition rate Ti:sapphire femtosecond laser. The beam is split with a polarizing beam splitter into a pump and a probe beams. The probe beam is introduced in an synchronously pumped OPO that allows to tune the wavelength and finally permit to do two-color pump-probe experiments (pump and probe are linearly and circularly polarized respectively). The transient optical reflectivity signals are obtained thanks to a mechanical stage delay (delay line) which enables a controlled arrival time of the probe beam regarding to the pump pulse excitation. The experiments were conducted with the front-front configuration with incident pump and probe beams perpendicular to the surface as shown in Fig. \ref{fig1}(c) and Fig. \ref{fig2}(a). In our experiments, the pump and probe wavelengths were fixed at 830 nm (1.495 eV) and 582 nm (2.13 eV) with a corresponding absorption length of $\sim$ 10.1 nm and $\sim$9.8 nm \cite{optBT}. This very small penetration depth is due to the specific electronic band structure in this energy range where interband transitions exist \cite{optBT}. In particular, in the energy range of around 1.2-2.2 eV, there is a sharp variation of both the real and imaginary partsof the dielectric constant which is a favorable situation for the detection of phonons \cite{yu,merlin}. Furthermore, in the experiments, the pump and probe are focused with a microscope objective providing a typical spot radius of $\approx$5 micrometers. Additional details of the setup can be found in previous recent articles \cite{ayouch,lejman}. 

\section{Results and discussion}

\begin{figure}[t!]
\centerline{\includegraphics[width=9cm]{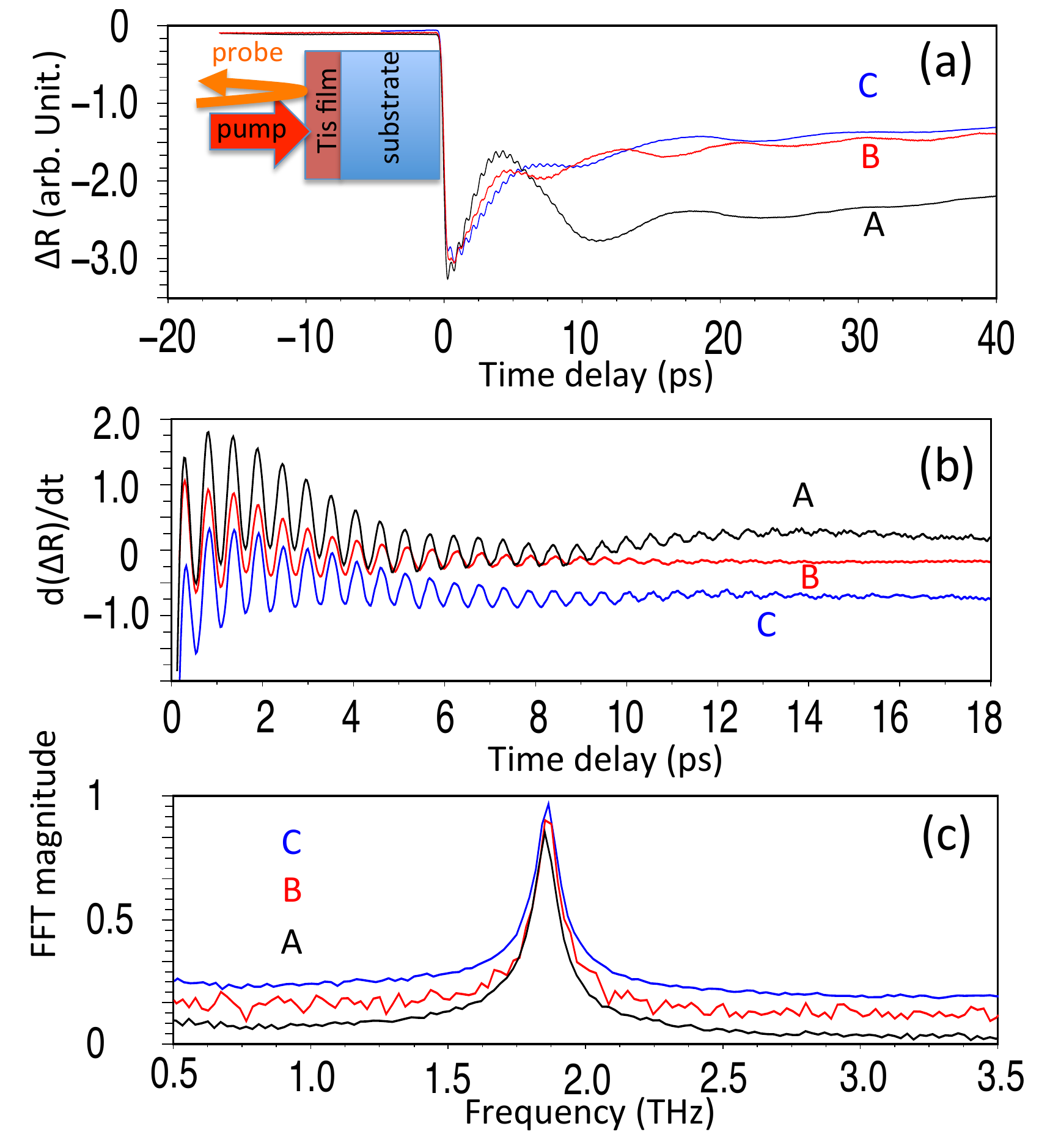}}
\caption{\label{fig2} (color online) (a) Time resolved optical reflectivity obtained for Bi$_2$Te$_3$ thin films deposited on various substrates: film A, B and C (see text for details about samples). The curves have been normalized to the largest magnitude point for clarity. (b) A1g(I) optical phonon signals. (c) corresponding Fast Fourier Transform (FFT) of curves shown in (b).}
\end{figure}

\begin{figure*}[t!]
\centerline{\includegraphics[width=18cm]{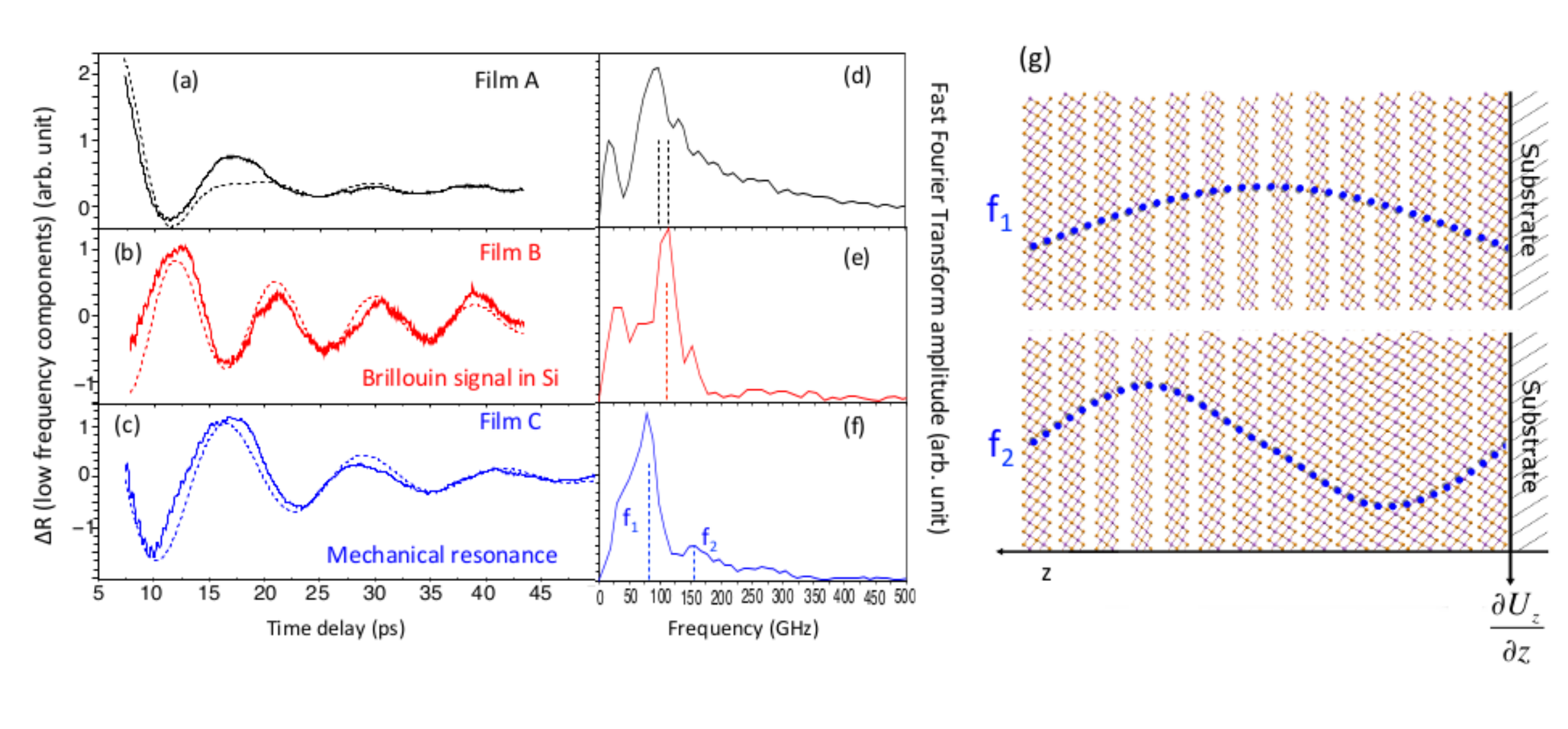}}
\caption{\label{fig3} (color online) (a)-(b)-(c) Coherent acoustic phonons signals extracted from the transient optical reflectivity signals. For clarity, the curves have been scaled in amplitude. The dotted lines are numerical adjustments (see the text for more details). (d)-(e)-(f) corresponding coherent acoustic phonons spectra obtained by a Fast Fourier Transform (FFT). (g) pictures of the two first confined acoustic eigenmodes in film C detected and whose frequencies are given in (f). U$_z$ is the normal displacement of atoms.}
\end{figure*}

The time resolved optical reflectivity obtained for the 3 samples are shown in Fig. \ref{fig2}(a). The signals are composed of different contributions. A first sharp variation of the optical reflectivity R corresponds to the electronic excitation by the pump light with a following decay of the out-of equilibrium carriers. Within the first picoseconds, several oscillatory components appear. The high frequency component, as shown in Fig. \ref{fig2}(b), is the signal of coherent longitudinal optical phonon (A1g(I)) well identified by its characteristic frequency  \cite{kama, nori,dekor,wang} and whose corresponding atoms displacements are shown on the structure displayed in Fig. \ref{fig1}(c). The Fast Fourier Transform (FFT) of this high frequency mode is shown in Fig. \ref{fig2}(c). The signals in Fig. \ref{fig2}(a) also show a rapid birth of low frequency components (coherent acoustic phonons spectra shown in Fig. \ref{fig3}(a)-(c)) even before the A1g(I) mode decays. 

These results, contrary to previous reports \cite{wang} and for similar range of photoexcitation, show that it is actually possible to generate and to detect the coherent optical phonons A1g(I) in BT nanostructures containing 10 QLs and no apparent restriction exists on the photoinduced forces for the investigated pump fluence range of 0.005-0.1 mJ.cm$^{-2}$. Despite we monitored the detection of the coherent phonons with a circular probe polarization, that integrates all the contributions from the transient dielectric tensor \cite{merlin,dress}, we did not detect the lower frequency transverse optical mode (Eg1(I)) neither the longitudinal optical mode A1g(II). Eg1(I) was neither observed in studies on bulk BT \cite{nori,kama,dekor,wang}. If we compare now the A1g(I) frequency (Fig. \ref{fig2}(c)), no detectable difference is reported between polycrystalline film, single crystalline film and crystallized islands. The phase of these modes are also very similar as shown in Fig. \ref{fig2}(b). No significant difference in the decay time appears indicating that the decay of the LO mode is not so sensitive to the long range crystallographic order in agreement, for example, with the observation of constant Raman linewidth (mode at 175 cm$^{-1}$) observed for exfoliated films having a variable nanostructures, i.e. a variable number of QLs of Bi$_{2}$Se$_{3}$ \cite{zhao}. This property indicates that the LO mode (A1g(I)) does not propagate in the film due likely to a very small group velocity and due to a rapid damping caused by the intrinsic anharmonic decay (phonon-phonon collision). This anharmonic coupling is known to be large in BT and leads to low thermal conductivity \cite{abinitBT,thermal}. 
We actually know how sensitive the optical phonons are to the unit cell forces distribution. As a matter of fact, this FFT features (same frequency than in bulk crystal) is thus a direct proof that the detected coherent optical phonon A1g(I) comes from the right BT structure with the right chemical environment in agreement with our XPS analysis. Consequently, the slight excess of Bi in film B and C and the excess of Te in film A might be isolated in the structure within probably interlayers as already suggested by XRR for the film A and probably within an interlayer between QL for Bi as suggested in the literature \cite{bos}. 

The generation of the A1g(I) mode even with ten QLs has to be discussed. Because of a large pump optical absorption in the case of our experiment, the relevant mechanism is not the impulsive stimulated Raman scattering (called IRSR or electrostriction) that does not require light absorption as stated years ago in the literature \cite{dress}. 
Furthermore, it is important to underline that despite the random distribution of the BT grains in the films A or the existence of crystallized islands (film B)  (see REED images in Fig. \ref{fig1}(a)), the generation/detection of the A1g(I) mode is still efficient and remarkable similarities are reported for the experimental properties of the A1g(I) mode. 
Then we believe that the mechanism is due to a large and local coupling between hot electrons and the A1g(I) mode. The mechanism of ultrafast generation of symmetry A1g modes has already been discussed in the literature \cite{dress} and is generally attributed to a hot electron-phonon coupling that preserves the lattice symmetry but the discussion of the quantitative generation mechanism requires the knowledge of the electron-phonon coupling parameter that usually lacks. In the particular case of BT, recent calculations have shown that the isotropic electron/optical phonon deformation potential is actually as large as d$_{e-op}$ $\approx$ 40eV \cite{abinitBT}. We have reported (Fig. \ref{fig2}) that for our nanostructures the LO A1g(I) mode has the same frequency than that of the bulk BT. This indicates that the interatomic potential are then similar in bulk BT and in our nanostructures, so that it is reasonable to use a bulk electron-optical deformation parameter, as a first approximation, to discuss the photoinduced forces. Then, following the convention used in semiconductors physics \cite{yu} (but different expressions of the electronic pressure containing the same physics also exist \cite{Bi1}), the associated photoinduced stress driven by deformation potential can be written as $ \sigma_{e-op}$=-d$_{e-op}$N 
where N is the photoexcited carriers concentration. Since  d$_{e-op}$ is positive this indicates first that in the early stage of excitation there is an expansion of the BT lattice. With a concentration of photoexcited carriers of N$\sim$ 0.5.10$^{27}$ m$^{-3}$ (typically that achieved with our maximum absorbed pump fluence of 0.1mJ.cm$^{-2}$) the photoinduced stress becomes as large as $\sigma_{e-op}$ $\sim$ -3 GPa. The force that induces atoms motion is defined as F$_{e-op}$ $\approx \partial{\sigma_{e-op}}/\partial{x}$. Because the pump skin depth is very small in BT ($\xi \approx$ 10 nm), this force becomes in magnitude F$_{e-op}$$\sim \sigma_{e-op}/ \xi$ = 4.10$^{17}$ J.m$^{-4}$ which is similar to what has been estimated for the A1g(I) mode in Bi for an equivalent range of pump excitation \cite{Bi1}. Contrary to what as been stated previously \cite{wang}, it is worth to underline that this force does not depend on the nanostructure thickness but rather on the gradient of the photoinduced stress. That is the reason why, on the basis of our observations and on our theoretical estimates, we believe this electron-optical deformation potential mechanism is likely the driving mechanism for coherent LO excitation. The photoinduced stress gradient can of course be smeared because of hot electrons diffusion, but we think that this diffusion should be somehow limited within a QL because of Van der Waals contacts. It is then interesting to discuss that LO excitation in that particular situation. If we consider, as a first approximation, that each QL are nearly isolated (or poorly connected, i.e. weak electron wavefunction overlapping) due to the Van der Waals bonds, then the photoexcited carriers can be initially mostly located within each QL. We can also consider as another approximation that the electrons are nearly homogeneously distributed within each QL (a rapid estimate shows that with a Fermi velocity of V$_F$ $\sim$ 3.6.10$^{5}$ m.s$^{-1}$ \ \cite{haj}, the distribution of hot electrons over a distance of 10 nm takes 30 fs). As a matter of fact, the photo-excitation induces an expansion of each QL which corresponds to a 1D-like breathing of each QL. This motion is in agreement with the symmetric atomic motion of the A1g(I) LO mode sketched with red arrows (Fig. \ref{fig1}(c)). 

The analysis of the photo-induced coherent acoustic phonons dynamics also provides some important informations on the elasticity of BT in confined geometries as well as on the electron-acoustic phonon coupling (Fig. \ref{fig3}(a)-(c)). The ultrafast light-induced coherent acoustic phonons are observed for the three samples with clearly different signals while no significant variation was observed for the LO mode A1g(I). 
In our experiments, with a characteristic penetration depth of pump light of $\sim$ 10 nm that typically scales with the film thickness, several QLs are nearly simultaneously excited by the femtosecond pump beam. This can lead to a collective and in phase motion (mechanical resonance) as already observed for similar experiments in different thin films \cite{ayouch,mechri}  or nanoparticles \cite{juve}. However, according to the elastic boundaries conditions of our system, we have to distinguish different situations. If the acoustic impedance Z=$\rho$V ($\rho$ is the mass density and V the sound velocity) of the film is close to that of the substrate then the acoustic phonons generated in the TI film are transmitted nearly perfectly into the substrate (no confinement of acoustic phonons, i.e. the acoustic reflection coefficient R$_{ac}$=(Z$_{film}$-Z$_{substrate}$)/(Z$_{film}$+Z$_{substrate}$) $<<$1 \cite{ayouch,mechri}). 
If the film and substrate impedances are different, then a part of the coherent acoustic phonons is confined leading to the out-of-plane ringing of the nanostructures \cite{ayouch,mechri}. Based on literature data known for bulk BT, it appears that the acoustic impedance (bulk longitudinal sound velocity V$_{BT}$=2600 m.s$^{-1}$ \ \cite{venSL}, $\rho_{BT}$=7460 kg.m$^{-3}$) is very close to that of (100) silicon (V$_{Si}$=8400 m.s$^{-1}$, $\rho_{Si}$=2200 kg.m$^{-3}$) but clearly larger than that of mica ($\rho_{mica}$=2790 kg.m$^{-3}$ and out-of-plane velocity V$_{mica}$=5027 m.s$^{-1}$ \ \cite{micaV}). As a consequence, for a perfect film/substrate interface, no acoustic confinement is expected for film deposited onto silicon. This situation is actually met for film B where most of the acoustic phonons energy is transmitted into the Si substrate (R$_{ac}$ $\sim$ 0.03). The emitted acoustic phonons are then detected in the substrate as evidenced by Brillouin mode in Si characterized by the frequency  f$_{Si}$=$2n_{Si}V_{Si}/\lambda$=115 GHz  in agreement with the observations (Fig. \ref{fig3}(e)), where $\lambda$=582 nm and $n_{Si}$=4\cite{optSi}. A numerical adjustment with a damped cosinus function gives $\Delta$R$_{Ac.}$(B) $\sim$ cos(2$\pi$f$_{B}$t$\times$+$\phi_{B}$)e$^{-t/\tau_B}$ with f$_{B}$=111 GHz and $\tau_{B}$=24 ps (dotted line in Fig. \ref{fig3}(b)). 
On the other hand, for the film C, the mode that is detected is not that of Brillouin nature in the mica substrate (a similar estimate shows that the Brillouin frequency is expected at f$_{mica}$=$2n_{mica}V_{mica}/\lambda$=27 GHz, with n$_{mica}$=1.6 \cite{micaopt}). The oscillatory components detected for film C actually correspond  to mechanical resonance eigenmodes due to the confinement of acoustic phonons (R$_{ac}$$\sim$0.17). The FFT shown in Fig. \ref{fig3}(f) evidences two modes (f$_{1}$$\approx$80 GHz and f$_{2}$$\approx$160 GHz) with f$_{2}$$\sim$ 2 f$_{1}$. This is the sequence of the expected harmonics of our resonator whose  frequency is given by  f$_{n}=nV_{C}/{2H_{C}}$ with n=1, 2, 3, ... and where V$_{C}$ is the longitudinal sound velocity of the film C and H$_{C}$ its thickness (Fig. \ref{fig3}(g)). A numerical adjustment with only the fundamental mode gives $\Delta$R$_{Ac.}$(C) $\sim$ cos(2$\pi$f$_1$t$\times$+$\phi_{C}$)e$^{-t/\tau_C}$, with $\tau_C$=14 ps (Fig. \ref{fig3}(c)). 
Thanks to this resonator model, we immediately obtain V$_{C}$=2460 m.s$^{-1}$ which is consistent with theoretical estimates in bulk BT ($\sim $ 2300 m.s$^{-1}$)\cite{yao} and with previous estimates obtained in pump-probe experiments carried on Bi$_{2}$Te$_{3}$/Sb$_{2}$Te$_{3}$ superlattice ($\sim $ 2600 m.s$^{-1}$) \cite{venSL}. Considering the mass density of the stoichiometric BT film $\rho$=7642 kg.m$^{-3}$, this leads to an elastic constant C$_{33}$=$\rho \times $V$^{2}$ $\approx$ 36-46 GPa also in accordance with recent calculations \cite{abinitBT}. Finally, according to this model, and because film A was grown on silicon too, we should observe only Brillouin mode in the silicon substrate. The spectrum in Fig. \ref{fig3}(d) shows a small Brillouin signal in Si with additional modes not resolved yet. An numerical adjustment (Fig. \ref{fig3}(a)) has been done with two damped cosinus (frequencies indicated by vertical dotted lines in Fig. \ref{fig3}(d)) and hardly reproduces the experimental signal. Because of complex interfaces and multilayers composition as well as polycrystalline texture (the relevant sound velocity is probably a mixture of out-of-plane and in-plane sound velocity), this spectrum requires further investigations which are not the topic of the present paper.  

Furthermore, in the case of film C, probing the assemblies of QLs with the coherent acoustic phonons could provide new insights on the QLs Van der Waals interactions. Usually in Van der Waals bonded solids \cite{ayouch} the macroscopic/mesoscopic elastic modulus is driven by the Van der Waals contacts, as a matter of fact if we consider it is the case for the 1D assemblies of QLs, then we obtain a Te-Te Van der Waals force elastic constant of the order of C$_{VdW}$ $\sim$ C$_{33}$$\times$ a$_{Te-Te}$ $\sim$  12-16 N.m$^{-1}$, where the Te-Te distance between two QL is a$_{Te-Te}$=0.364 nm\cite{strucBT}. These estimates are 1.5 to 2 times larger than previous estimates obtained thanks to macroscopic estimate \cite{strucBT}. As noticed previously, we have to keep in mind that these constants C$_{VdW}$ remains ten times larger than real Van der Waals interactions existing in rare gas solids \cite{abinitBT}.

Finally, it is necessary now to discuss the physical origin of the acoustic phonon emission. A rapid estimate shows that the driving mechanism of coherent acoustic phonon generation is of electronic nature. The evaluation of the photoinduced thermoelastic stress ($\sigma_{Th}$), due to a rapid lattice heating that follows the electronic decay, can be accounted following the standard model \cite{tom1}. 
We find $\sigma_{Th}=-3\beta B \Delta T$  $\sim$ 0.1-0.2 GPa with B $\approx$ 37 GPa the bulk modulus \cite{abinitBT}, $\beta$=2.10$^{-5}$K$^{-1}$ \ \cite{betaperpen} the out-of-plane BT thermal expansion and $\Delta$T=$\Delta$E/C$_{L}$ $\approx$ 60 K  the lattice heating where C$_{L}$=1.5.10$^{6}$ J.m$^{-3}$.K$^{-1}$ \ \cite{CLBT} is the BT lattice heat capacity and $\Delta$E=N$\times$ 1.495 eV$\sim$10$^{8}$ J.m$^{-3}$ the total laser pump energy per unit of volume absorbed by the material (this maximum temperature increase is obtained with a fluence of 0.1 mJ.cm$^{-2}$ and without considering the heat conductivity at film/substrate interface). This thermoelastic stress is much smaller than the electron-acoustic phonon deformation potential stress given by \cite{tom1,ruello2009} $\sigma_{e-ac} \approx -a_{e-ac}N  $=-2.5 GPa, were a$_{e-ac}$=35 eV \cite{abinitBT} is the electron/acoustic deformation potential (four times larger than that in GaAs \cite{yu}) and N$\sim$ 0.5.10$^{27}$ m$^{-3}$. The prevailing contribution of the electron-acoustic phonon deformation potential mechanism is also consistent with the evaluation of this scattering process made on the basis of recent transport properties measurements \cite{kim}.


\section{Summary}

As a summary, thanks to a complete ultrafast two-colors pump-probe investigation performed on various films grown by state-of-art MBE, we show that it is possible to excite coherent A1g(I) LO mode in thin films containing ten QLs and no restriction appears. The A1g(I) LO phonon dynamics is nearly not disturbed by the nanostructure of the films which indicates that the decay takes place locally by anharmonic coupling rather than by scattering due to geometrical inhomogeneities like islands (film B) or grain boundaries (film A) . Our estimates show that the LO deformation potential photoinduced stress appears to be as high as -3 GPa for a photoexcited carriers concentration of N$\sim$ 0.5.10$^{27}$ m$^{-3}$ which is quite large and likely to launch the atomic motion of A1g(I) mode in our nanostructures. It is important to notice here that we did not detect two different A1g(I) mode components arising from the so-called bulk and surface phonons as it has been reported recently on bulk Bi$_{2}$Se$_{3}$ with time-resolved ARPES (Angle Resolved PhotoEmission Spectroscopy) \cite{sobota}. The origin of the difference is not clear yet, since this could be either due to the confinement of our nanostructures or due to the lack of sensitivity of optical methods even in the particular case of BT, the pump/probe light does penetrate only around 10 nm which is still a characteristic distance over which surface and bulk electrons may interact \cite{haj}. This point will need to be clarified in the near future. Beside the longitudinal optical phonon dynamics, we also show, before the LO mode decays, the birth of coherent acoustic phonons whose spectrum drastically depends on the films nanostructure contrary to that of the LO mode. This clearly show the sensitivity of the LA modes to the nanostructure arrangement. The generation mechanism of these acoustic phonons is also attributed to the deformation potential mechanism. Finally, the measurement of the time of flight (sound velocity) of these longitudinal acoustic phonons in the single crystalline film (C) provides an evaluation of the out-of-plane elastic modulus (36-46 GPa) of these assemblies of QLs. As a final perspective, the coexistence of both optical and acoustic phonons could make these systems as a rich playground for studying the LO-LA anharmonic couplings which are two important phonons populations involved in the thermal properties  \cite{abinitBT,thermal}.

Acknowledgements : we thank G. Vaudel, V. Gusev and A. Bulou for helpful discussions. This work was supported by the French Ministry of Education and Research, the CNRS and Region Pays de la Loire (CPER Femtosecond Spectroscopy equipment program). R. Rapacz was supported by FORSZT PhD fellowship.

\newpage

\end{document}